\documentclass[12pt, prd, showpacs]{revtex4}
\usepackage{amssymb}
\usepackage{amsmath}

\input{tcilatex}

\begin{document}

\title{Super-Penrose process for extremal charged white holes}
\author{O. B. Zaslavskii}
\affiliation{Department of Physics and Technology, Kharkov V.N. Karazin National
University, 4 Svoboda Square, Kharkov 61022, Ukraine}
\affiliation{Institute of Mathematics and Mechanics, Kazan Federal University, 18
Kremlyovskaya St., Kazan 420008, Russia}
\email{zaslav@ukr.net }

\begin{abstract}
We consider collision of two particles 1 and 2 near the horizon of the
extremal Reissner-Nordstr\"{o}m (RN) black hole that produce two other
particles 3 and 4. There exists such a scenario that both new particles fall
in a black hole. One of them emerges from the white hole horizon in the
asymptotically flat region, the other one oscillates between turning points.
However, the unbounded energies $E$ at infinity (super-Penrose process -
SPP) turn out to be impossible for any finite angular momenta $L_{3.4}$. In
this sense, the situation for such a white hole scenarios is opposite to the
black hole ones, where the SPP is found earlier to be possible for the RN
metric even for all $L_{i}=0$. However, if $L_{3,4}$ themselves are
unbounded, the SPP does exist for white holes.
\end{abstract}

\keywords{particle collision, super-Penrose process, white holes}
\pacs{04.70.Bw, 97.60.Lf }
\maketitle

\section{Introduction}

In last decade, great interest is provoked in high energy collisions of
particles in strong gravitational field. This happened after findings in 
\cite{ban}, where it was shown that the energy in the center of mass $%
E_{c.m.}$ can grow unbounded if collision takes place near the horizon of
the extremal Kerr black hole (this is called the BSW effect, after the name
of its authors). Later, many papers appeared in which a list of potential
sources of high energy collisions was extended. In particular, it includes
white holes. One of scenarios consists in collision in our universe between
particles one of which falls into a black hole while the second one emergs
from a white hole \cite{tot}, \cite{white}. Recently, another scenario was
considered for the extremal Kerr metric in which collision happens near the
black hole but, afterwards, particles cross the horizon, leave our universe
and appear in a white hole region \cite{ph}. As a result, high energy fluxes
can be registered in another universe. Or, vice versa, if the process
started in some another world, this leads to high energy particles that are,
in principle, can be detected in our Universe, so this would possibly give
explanation of astrophysically relevant high energy processes.

Our aim is to consider a charged counterpart of \cite{ph}. By definition, if 
$E$ is unbounded, one can speak about the super-Penrose process (SPP).
Instead of rotating black holes, we consider a more simple case of the
Reissner-Nordstr\"{o}m (RN) black-white hole. It was shown earlier that in
such a metric, the counterpart of the effect found in \cite{ban} also occurs 
\cite{jl}. Moreover, it turned out that for the RN metric the SPP\ is also
possible \cite{rn} in contrast to the case of rotating black holes \cite{is}%
. Thus there are two mutually complimentary cases, each of which deserves
special attention - rotating and static charged black-white holes. In the
present case we consider just the second option. We elucidate, whether or
not the SPP is possible for such white holes.

Some reservations and additional arguments concerning our motivation are in
order. There are strong factors that testify in favour of the instability of
white holes \cite{unst} (see also Sec. 15 of \cite{fn}). However, the
nontrivial structure of the RN space-time that includes white hole regions 
\cite{bc} follows from the theory anyway, so the complete theory of the BSW
effect should include in consideration the corresponding version of this
effect. Moreover, white holes can reveal themselves as windows from other
worlds through which energy can flow into ours \cite{nov}, \cite{dad}. In
this context, a new potential source of ultra high energy can be just one
more manifestations of white holes instability thus deserving to be studied.
Also, we would like to draw attention to the following detail. After the
paper \cite{ban} the interest to more earlier works on high energy
collisions near the horizon was revived \cite{katz}, \cite{ps}. Meanwhile,
the head-on collisions considered in \cite{ps} (see eq. 2.57 there)
correspond just to white holes.

\section{Basic equations}

Let us consider the metric%
\begin{equation}
ds^{2}=-dt^{2}N^{2}+\frac{dr^{2}}{N^{2}}+r^{2}d\omega ^{2}\text{,}
\end{equation}%
where $d\omega ^{2}=d\theta ^{2}+\sin ^{2}\theta d\phi ^{2}\,.$ We will deal
with the extremal RN metric for which%
\begin{equation}
N=1-\frac{r_{+}}{r}\text{,}
\end{equation}%
where $r_{+}=Q=M$ is the horizon radius, $Q$ being the electric charge of a
black hole, $M$ its mass. We use the system of units in which fundamental
constants $G=c=1$.

For a particle with the electric charge $q$ and the mass $m$ the equations
of motion within the plane $\theta =\frac{\pi }{2}$ read%
\begin{equation}
m\dot{t}=\frac{X}{N^{2}}\text{,}
\end{equation}%
\begin{equation}
m\dot{r}=\sigma P\,\text{, }\sigma =\pm 1,
\end{equation}%
\begin{equation}
P\equiv \sqrt{X^{2}-\tilde{m}^{2}N^{2}}\text{, }\tilde{m}^{2}\equiv m^{2}+%
\frac{L^{2}}{r^{2}}\text{,}  \label{P}
\end{equation}%
\begin{equation}
m\dot{\phi}=\frac{L}{r^{2}}\text{.}
\end{equation}%
Here%
\begin{equation}
X=E-q\varphi \text{,}
\end{equation}%
$E$ is the energy, $\varphi =\frac{r_{+}}{r}=1-N$ is the electric potential
of the extremal RN black hole, $L$ being the angular momentum, $\sigma =\pm
1,\,$dot denotes derivative with respect to the proper time $\tau $. The
forward-in-time condition requires 
\begin{equation}
X\geq 0.  \label{ft}
\end{equation}

Let us consider the following scenario. Particles 1 and 2 fall from
infinity, collide in point $r=r_{c}$ close to the horizon and create
particles 3 and 4. Thus $\sigma _{1}=\sigma _{2}=-1$. The conservations of
the electric charge, energy and radial momentum in the point of collision
give us%
\begin{equation}
X_{0}\equiv X_{1}+X_{2}=X_{3}+X_{4}\text{,}  \label{x12}
\end{equation}%
\begin{equation}
q_{0}\equiv q_{1}+q_{2}=q_{3}+q_{4}\text{,}
\end{equation}%
\begin{equation}
L_{0}\equiv L_{1}+L_{2}=L_{3}+L_{4}\text{,}  \label{L}
\end{equation}%
\begin{equation}
E_{0}\equiv E_{1}+E_{2}=E_{3}+E_{4},  \label{e}
\end{equation}%
\begin{equation}
-P_{1}-P_{2}=\sigma _{3}P_{3}+\sigma _{4}P_{4}\text{.}  \label{p12}
\end{equation}

We will use the standard classification. If $X_{H}=0$ (subscript "H" means
that a corresponding quantity is taken on the horizon), a particle is called
critical. If $X_{H}\neq 0$ is separated from zero, it is called usual. If $%
X_{H}\neq 0$ but is very small (of the order $N_{c}\equiv N(r_{c})$), it is
called near-critical. As for the extremal RN black hole, $\varphi (r_{+})=1$%
, the criticality condition reads $E=q$.

For the critical particle,%
\begin{equation}
X=EN\text{, }  \label{crit}
\end{equation}%
\begin{equation}
P=N\sqrt{E^{2}-\tilde{m}^{2}}\text{.}  \label{pcrit}
\end{equation}

If $N\ll 1$, we have for the usual particle,%
\begin{equation}
X=E-q+qN\text{,}
\end{equation}%
\begin{equation}
P=X+O(N^{2})\text{.}  \label{pu}
\end{equation}

In what follows, we are interested in high-energy processes, so we assume
that particle 1 is critical and particle 2 is usual. This choice guarantees
that the energy in the center of mass frame $E_{c.m.}$ is unbounded \cite{jl}%
.

\section{Types of scenario}

We consider collision near the horizon, so $r_{c}\approx r_{+}$. Assuming
that all masses and angular momenta are finite and taking the limit $%
r_{c}\rightarrow r_{+}$, one can infer from (\ref{x12}) and (\ref{p12}) that
particles 3 and 4 cannot be both usual. Let particle 3 be near-critical and
particle 4 be usual. It is convenient to write for a near-critical particle 
\begin{equation}
q=E(1+\delta )\text{,}
\end{equation}%
where $\delta \ll 1$. As it is substantiated in \cite{rn}, it makes sense to
take $\delta $ of the order $N_{c}$, so%
\begin{equation}
\delta =C_{1}N_{c}+C_{2}N_{c}^{2}+...\text{.}  \label{d}
\end{equation}%
Actually, the terms of $N_{c}^{2}$ and higher can be neglected. Then, we
have for such a particle

\begin{equation}
X=NE(1-C_{1})+O(N^{2})\text{,}  \label{xnc}
\end{equation}%
\begin{equation}
P=N\sqrt{E^{2}(1-C_{1})^{2}-\tilde{m}^{2}}+O(N^{2})\text{.}  \label{ncr}
\end{equation}

Then, we can classify the scenarios of collision by means of two parameters.
If immediately after collision a particle moves inward, the scenario is
called IN, if it moves outward, the scenario is called OUT. And, depending
on the sign of $C_{1}$, we write $+$ or $-$. As a result, we have 4 possible
cases OUT$+$,OUT$-$, IN$+$, IN$-$ . The first three of them were already
analyzed in \cite{rn}. What remains to be seen is the property of scenario IN%
$-$. It was rejected in \cite{rn} since it corresponds to fall of both
particles into a black hole. However, it is this scenario that is of
interest to us now since it ensures the energy transfer to a white hole
region (see details below).

For our scenario IN$-$ we have $\sigma _{3}=-1$. Then, it follows from (\ref%
{x12}) and (\ref{p12}) for $r_{c}$ close to $r_{+}$ that $\sigma _{4}=-1$.
In other words, two particle collide near the black hole horizon and enter
the inner region. Now, we are going to elucidate, whether or not in this
process $E_{3}>0$ can be unbounded. If yes, $E_{4}=E_{0}-E_{3}$ is negative
and unbounded for any finite $E_{0}$ in (\ref{e}). The properties of
corresponding trajectories are described below.

\section{Dynamics of collision}

It follows from (\ref{crit}) - (\ref{pu}) and (\ref{xnc}), (\ref{ncr}) that
the conservation of the radial momentum (\ref{p12}) with $\sigma _{3}=\sigma
_{4}=-1$ can be rewritten in the form similar to that used in \cite{rn}:%
\begin{equation}
F=-\sqrt{E_{3}^{2}(1-C_{1})^{2}-\tilde{m}_{3}^{2}},  \label{Fn}
\end{equation}%
where 
\begin{equation}
F\equiv A+E_{3}(C_{1}-1)\text{,}
\end{equation}%
\begin{equation}
A\equiv E_{1}-\sqrt{E_{1}^{2}-\tilde{m}_{1}^{2}}\text{.}  \label{a}
\end{equation}

Taking the square of (\ref{Fn}), we obtain%
\begin{equation}
C_{1}=1-\frac{\tilde{m}_{3}^{2}+A^{2}}{2AE_{3}}\text{,}
\end{equation}%
\begin{equation}
F=\frac{A^{2}-\tilde{m}_{3}^{2}}{2A}\text{.}
\end{equation}%
As, for our scenario IN$-$, $C\ _{1}<0$, we immediately obtain that%
\begin{equation}
E_{3}<\frac{\tilde{m}_{3}^{2}+A^{2}}{2A}\text{.}  \label{lb}
\end{equation}

We see that $E_{3}$ is bounded from the above, so SPP in the white hole
region is impossible. From another hand, the condition $F<0$ that follows
from (\ref{Fn}), gives us a lower bound on the effective mass, $\tilde{m}%
_{3}>A$.

The result about impossibility of the SPP retains its validity if, instead
of the given process we consider its Schnittmann analogue \cite{sch}, when
the critical particle 1 comes from the horizon, so $\sigma _{1}=+1$. The
only changes is that $A=E_{1}+\sqrt{E_{1}^{2}-m_{1}^{2}}$ instead of (\ref{a}%
).

\section{Rapidly rotating particles}

In the above consideration, it was assumed that $L_{3}$ is bounded. Then,
because of finiteness of the total angular momentum $L_{0}$ (\ref{L}), the
quantity $L_{4}$ is bounded as well. Meanwhile, there is a separate
question: is it possible to achieve large $E_{3}$ due to large $L_{3}$? If
yes, restriction (\ref{lb}) becomes irrelevant. For a scenario of such a
type, one has to take into account large $L_{3},_{4}$ from the very
beginning, already in $P_{3,4}$. A new picture, qualitatively different from
the one considered above, arises if%
\begin{equation}
L_{3,}=\frac{l_{3}}{\sqrt{N_{c}}}\text{, }L_{4}=L_{0}-\frac{l_{3}}{\sqrt{%
N_{c}}}\text{,}  \label{l34}
\end{equation}%
where $l_{3}$ does not contain small parameters. In this case, the analysis
of eq. (\ref{p12}) should be carried out anew. Then, taking the limit of $%
N_{c}\rightarrow 0$ and equating the terms of the zeroth order with respect
to $N_{c}$ in eq. (\ref{p12}), we obtain that $\sigma _{3}=\sigma _{4}=-1$.
For particle 1we can use (\ref{crit}), (\ref{pcrit}), for particle 2 it is
sufficient to take (\ref{pu}). For particles 3, 4 the centrifugal terms with 
$L_{3,4}^{2}$ in $P_{3,4}$ give the correction that shoud be taken into
account:%
\begin{equation}
P_{3,4}\approx \sqrt{X_{3,4}^{2}-N_{c}\frac{l_{3,4}^{2}}{r_{+}^{2}}}\approx
X_{3}-\frac{N_{c}}{2X_{3}}\frac{l_{3,4}^{2}}{r_{+}^{2}}\text{. }
\end{equation}

Collecting all terms of the order $N_{c}$ and, one can obtain from the
conservation laws (\ref{x12}), (\ref{p12}) that%
\begin{equation}
A=\frac{l_{3}^{2}}{2r_{+}^{2}}(\frac{1}{X_{3}}+\frac{1}{X_{4}})\text{.}
\end{equation}%
Taking into account (\ref{x12}) one more time, we obtain the final
expression 
\begin{equation}
\left( X_{3,4}\right) _{c}=\frac{X_{0}}{2}(1\pm \sqrt{1-b}),  \label{xc}
\end{equation}%
\begin{equation}
b=\frac{2l_{3}^{2}}{r_{+}^{2}X_{0}A}\text{.}
\end{equation}%
It is implied that $b<1$. Then,%
\begin{equation}
E_{3}=\left( X_{3}\right) _{c}+q_{3}\varphi (r_{c})\approx \left(
X_{3}\right) _{c}+q_{3}.  \label{eq}
\end{equation}

In doing so, there is no bound like (\ref{lb}) at all. This is because both
particles 3 and 4 are usual, so the conditions $P_{3,4}^{2}>0$ are satisfied
automatically since $X=O(1)$ and $N_{c}L=O(\sqrt{N_{c}})\rightarrow 0$ in (%
\ref{P}) and there are no additional constraints. Both particles fall in a
black hole. Thus we can have formally unbounded $E_{3}$ provided $q_{3}$ is
also unbounded, to keep $X_{3}$ finite. Actually, there are no unbounded $q$
in nature ($\left\vert q\right\vert <Z_{e}\left\vert e\right\vert $, where $%
Z_{e}\approx 170,$ $e$ being the electron charge) that restricts the value
of $E_{3}$ in a way similarly to what takes place in black hole scenarios 
\cite{rn} (see also Sec. V in \cite{esc} for discussion of macroscopic
charged bodies). But $E_{3}$ is sufficiently large anyway, according to (\ref%
{eq}).

\subsection{Trajectories with $E>0$ beyond a black hole horizon}

The expressions (\ref{xc}) are valid near the point of collision. To gain
some energy due to particle 3 in the asymptotically flat region in the white
hole zone, we need (i) the turning point that prevents a particle from
falling in the singularity, (ii) the absence of the turning point outside
the next horizon, for $r>r_{+}\,.$ To simplify formulas, let us consider the
case when $m_{3}=0$ or is negligible. Condition (i) is satisfied
automatically, if $E_{3}>q_{3}$ that is indeed valid according to (\ref{eq}%
). Then, the location of the turning point $r_{0}<r_{+}$ is%
\begin{equation}
\frac{r_{+}}{r_{0}}=\frac{1}{2}(1-\frac{q_{3}r_{+}}{L_{3}})+\sqrt{\frac{1}{4}%
(1-\frac{q_{3}r_{+}}{L_{3}})^{2}+\frac{Er_{+}}{L_{3}}}>1\text{.}
\end{equation}

Condition (ii) is satisfied, if 
\begin{equation}
q_{3}>\frac{L_{3}}{r_{+}}.  \label{qL}
\end{equation}
If $L_{3}$ obeys (\ref{l34}), we can take%
\begin{equation}
q_{3}=\frac{\alpha }{r_{+}\sqrt{N_{c}}}\text{, }
\end{equation}%
with%
\begin{equation}
\alpha >l_{3}\text{.}
\end{equation}

Then,%
\begin{equation}
E_{3}\approx \left( X_{3}\right) _{c}+\frac{\alpha }{r_{+}\sqrt{N_{c}}}
\end{equation}%
can be made as large as we like due to sufficiently small $N_{c}$. Thus the
SPP does exist.

\subsection{Trajectories with $E<0$}

Particle 3 in the scenario under discussion has large $E_{3}>0$, so particle
4 has large $E_{4}<0$. It follows from (\ref{ft}) that now $%
q_{4}=-\left\vert q_{4}\right\vert $, $X_{4}=\left\vert q_{4}\right\vert 
\frac{r_{+}}{r}-\left\vert E_{4}\right\vert $. It is clear that such a
particle cannot escape to infinity since this would violate (\ref{ft}). It
oscillates between turning points. They can be found from the condition $%
P_{4}=0.$ If $m_{4}=0$ or is negligible, the corresponding equation is
solved in a compact form,%
\begin{equation}
\frac{r_{+}}{\left( r_{0}\right) _{out}}=\frac{1}{2}(1-\frac{\left\vert
q\right\vert r_{+}}{\left\vert L\right\vert })+\sqrt{\frac{1}{4}(1-\frac{%
\left\vert q\right\vert r_{+}}{\left\vert L\right\vert })^{2}+\frac{%
\left\vert E\right\vert r_{+}}{\left\vert L\right\vert }}
\end{equation}%
outside the horizon, $\left( r_{0}\right) _{out}>r_{+}$. For shortness, we
omit subscript "4".

The generalized ergoregion \cite{ruf} lies at $E=0$, (\ref{qL}) for particle
4, and small but nonzero mass,%
\begin{equation}
\frac{r_{erg}}{r_{+}}\approx \frac{\sqrt{q^{2}-\frac{L_{3}}{r_{+}^{2}}^{2}}}{%
m}\gg 1\text{,}
\end{equation}%
We have taken into account that $L_{4}=L_{0}-L_{3}\approx -L_{3}$ and eq. (%
\ref{qL}).

Thus $\left( r_{0}\right) _{in}<r_{erg}$ and the turning point lies inside
the ergoregion, as it should be. There is no coincidence with eq. (12) of 
\cite{ruf} since it corresponds to $L=0$, whereas we consider the opposite
case $m\ll \frac{\left\vert L\right\vert }{r_{+}}$.

Inside the black hole horizon

\begin{equation}
\frac{r_{+}}{\left( r_{0}\right) _{in}}=\frac{1}{2}(1+\frac{\left\vert
q\right\vert r_{+}}{\left\vert L\right\vert })+\sqrt{\frac{1}{4}(1+\frac{%
\left\vert q\right\vert r_{+}}{\left\vert L\right\vert })^{2}-\frac{%
\left\vert E\right\vert r_{+}}{\left\vert L\right\vert }}.
\end{equation}

One can check that the conditions (\ref{qL}) and $\left\vert
E_{4}\right\vert =-E_{4}=\left\vert q_{4}\right\vert -X_{4}<\left\vert
q_{4}\right\vert $ do guarantee that $\left( r_{0}\right) _{out}>r_{+}$, $%
\left( r_{0}\right) _{in}<r_{+}$ , so the picture is self-consistent.

One can also introduce the notion of the ergoregion inside the horizon,
there the situation is opposite, $r_{erg}<\left( r_{0}\right) _{in}<r_{+}$
where now%
\begin{equation}
\frac{r_{+}}{r_{erg}}=1+\frac{\left\vert q\right\vert r_{+}}{\left\vert
L\right\vert }\text{.}
\end{equation}

Thus the particle in question crosses the black hole horizon $r_{+}$, enters
the white hole region, bounces in the turning point $\left( r_{0}\right)
_{in}$ and moves to larger radii, crosses the new horizon $r_{+},$ bounces
in the point $\left( r_{0}\right) _{out}$, falls inside the horizon $r_{+}$
again, etc. Earlier, it was pointed out in \cite{gpneg} that in the Kerr
metric a particle with $E<0$ cannot remain in the outer region and
necessarily dives inside the horizon, where it either falls in a singularity
or extends to an infinite affine distance, remaining inside the original
horizon. We see that both cases with particles with negative energies are
similar in this sense.

\section{Summary of results}

Thus we showed that particle collision on our side on the horizon do not
lead to unbounded energies in the white hole region, if all parameters of
original particles (masses, angular momenta, charges) are finite. In this
sense, for the RN metric the black and white hole scenarios under discussion
are complementary to each other. There exists the SPP for the black hole
case \cite{rn} but there is no such a process for the white hole case. From
another hand, there are special subcases, when a particles created in
collision have unbounded angular momenta. For them, the SPP does indeed
exist. In doing so, the electric charge should be also large (formally
unbounded). Thus white holes can be indeed sources of ultrahigh energy
fluxes in our universe created in the other ones but with reservation that
the corresponding matter or radiation should be rapidly rotating.

\section{Special point: SPP versus mass inflation}

In our approach, we did not take into account the process of mass inflation
that can change seriously the interior of a charged black hole as compared
to the "pure" RN solution \cite{pi} (see also Sec. 14 of \cite{fn} and
references therein). In view of this, one would think that the present work
has at least methodical character in the sense that it fills some previous
gaps in the theory of the SPP. \ It tends to the goal of searching this
phenomenon as fully as possible. Meanwhile, there are additional arguments
why our results can have not only pure methodical character but give
something more. In principle, the mass inflation that develops due to wave
propagation inside a black hole, on one hand, and collisions of massive
particles produced with unbounded energy, from the other had, can be thought
of as two different mechanisms whose backreaction can have drastic
consequences on a black hole geometry. How mass inflation and the SPP inside
a black hole can interfere is a very complicated process. In this context,
it makes reasonable to look, in the first approximation, at the possibility
of SPP alone, neglecting all other potential factors. And, as our results
show, the SPP is, typically, absent. This means that, in general, the mass
inflation is expected not to be affected seriously by collisions of massive
particles. There is a special situation, however, in which created particles
have unbounded angular momenta. Whether and how this can interfere with mass
inflation remains unclear and requires further investigations. 

Anyway, we would like to stress that the conclusion about the
presence/absence of the SPP are obtained by us by considering the
inequalities that describe the energies of particles that enter the horizon.
They are obtained by the limiting transition from the outer region where the
mass inflation is irrelevant at all. In this sense, whatever events would
happen inside and how the metric would look like, does not affect our main
conclusions about the presence/absence of the upper bound on $E$ for
particles that enter the horizon from the outside. 

\begin{acknowledgments}
This work was supported by the Russian Government Program of Competitive
Growth of Kazan Federal University.
\end{acknowledgments}

\end{document}